\documentclass[aps,prl,twocolumn,groupedaddress]{revtex4}

\usepackage{graphicx}
\usepackage{dcolumn}
\usepackage{bm}
\usepackage{color}

\begin{document}


\title{Discontinuous Lifshitz Transition Achieved by Band-Filling Control in Na$_x$CoO$_2$
}

%

\author{Yoshihiko Okamoto, Atsushi Nishio, and Zenji Hiroi}
\affiliation{
Institute for Solid State Physics, University of Tokyo, Kashiwanoha 5-1-5, Kashiwa 277-8581, Japan\\
}

\date{\today}

\begin{abstract}
Precise band-filling control is performed on a strongly correlated electron system Na$_x$CoO$_2$, by changing the Na content, and the influence of this on the magnetic and thermodynamic properties is studied.
It has been discovered that a phase boundary between a Pauli paramagnetic metal appearing at small $x$ and a Curie-Weiss metal at large $x$ is located within an extremely narrow range of
 0.620 $<$ $x$ $<$ 0.621.
The transition between these is a Lifshitz transition of the pocket-vanishing type, exceptionally achieved by band-filling control. 
Abrupt changes in various properties
across the boundary are remarkably observed, suggesting that the Lifshitz transition occurs discontinuously in Na$_x$CoO$_2$, differently from the conventional Lifshitz transition defined at $T$ = 0.
\end{abstract}

\maketitle

Half a century ago, Lifshitz considered an electronic phase transition associated with a change of Fermi surface (FS) topology induced by continuous lattice deformation for noninteracting fermions~\cite{1}. This transition, now called the Lifshitz transition (LT), is unique in the sense that it is not associated with symmetry breaking, which is always expected for most phase transitions of the Landau type. Moreover, it is assumed to be a continuous transition defined only at $T$ = 0 and becomes a crossover at finite temperatures. Lifshitz examined two types of changes in the FS topology: one is associated with the collapse of a cylindrical FS, and the other is characterized by the emergence of an additional FS~\cite{1}. These are classified as the neck-collapsing type and the pocket-vanishing type respectively. As a control parameter for these transitions, he assumed a lattice deformation under high pressure, while pointed out that band-filling control by chemical substitution would be unsuitable, because it inevitably brought disorder into the system.

Many experiments have been carried out to elucidate the nature of the LT. In the initial stage, nonlinear pressure dependence of superconducting transition temperatures in Tl and Re was explained by the LT~\cite{3,4}. Later, several compounds have been assumed to generate pressure-induced LTs, e.g., a crossover for intermetallic alloys like AuIn$_2$~\cite{5} and metamagnetic transitions in itinerant ferromagnets, UGe$_2$ and ZrZn$_2$~\cite{6,7}. 
In most cases, however, effects due purely to changes in the FS topology are small and sometimes obscured by thermal effects and complicated FSs. It seems that, only when other degrees of freedom like lattice or spin couple strongly with electronic states, the transition manifests itself dramatically. 

We focus on Na$_x$CoO$_2$ as a candidate compound to generate the LT by band-filling control.
Na$_x$CoO$_2$ is a quasi two-dimensional metal and well known to show a large thermoelectric power at $x$ $\sim$ 0.7 or superconductivity by hydration~\cite{8,9,10}. As shown in Fig. 1, it crystallizes into a layered structure made up of alternately stacked CoO$_2$ and Na$_x$ layers, which serve as conducting and charge-reservoir layers respectively~\cite{10}. Na$_x$CoO$_2$ is synthesized 
in a wide range of 0.1 $\le$ $x$ $\le$ 1.0 by a soft chemistry technique~\cite{11,12}. 
It should be noted that the band filling is generally controlled by this large Na nonstoichiometry.
In most cases, Na ions tend to be randomly distributed in the Na layer and, therefore, have little influence on the conducting layers, resulting in a nearly ideal band-filling control. However, there are a few exceptional cases where the Na ions align periodically at certain fractional values of $x$, disturbing the electronic state of the CoO$_2$ layer and even leading to a charge-ordered insulating ground state at $x$ = 0.5~\cite{13}.

The $x$-dependence of physical properties of Na$_x$CoO$_2$ has been investigated by several groups~\cite{13,14,15,Lang}. They show that a Na-rich phase ($x$ $\sim$ 0.7) is a Curie-Weiss (CW) paramagnetic metal and a Na-poor phase ($x$ $\sim$ 0.3) is a Pauli paramagnetic (PP) metal. Although there must be a phase boundary between them, no consensus has been obtained on the boundary value, referred to as $x$*. For example, Foo \textit{et al}. reported $x$* $\sim$ 0.5~\cite{13}, while Yokoi \textit{et al}. reported $x$* $\sim$ 0.6~\cite{14}; they used a series of samples prepared by the soft chemistry method. In contrast, Yoshizumi \textit{et al}. succeeded in controlling the Na content precisely in the range of 0.3 $\le$ $x$ $\le$ 0.7 by a solid-state reaction instead of the soft chemistry method and found that the phase boundary is located at $x$* $\sim$ 0.58 $\pm$ 0.01~\cite{15}. This inconsistency may be due to the tendency towards Na ordering near $x$ = 0.5 as well as the difficulty in finely tuning $x$. Hence, it is important to locate the boundary more systematically using well-characterized samples and to study in detail how physical properties change near the boundary as a function of $x$. 

Band-structure calculations on Na$_x$CoO$_2$ show that a $t_{\mathrm{2g}}$ band from the Co 3$d$ orbitals is responsible for most electronic properties; this band splits into an $a_{\mathrm{1g}}$ band and a doubly degenerate $e_{\mathrm{g}}^{\prime}$ band~\cite{16,17,18}.
The $a_{\mathrm{1g}}$ band always forms a large hole FS, irrespective of $x$, with a cylindrical shape reflecting the two dimensional crystal structure. 
In addition, a small electron pocket 
possibly appears at the $\Gamma$ point.
Although this pocket has 
remained undetectable in angle-resolved photo-emission-spectroscopy experiments~\cite{19}, it must be present, at least at large $x$ values according to most of the band-structure calculations~\cite{16,17,18}. Theory predicts that this electron pocket should give rise to two kinds of characteristic fluctuation: an extended $s$-wave superconducting fluctuation when it is small, and a ferromagnetic spin fluctuation when it becomes moderately large~\cite{18}. In fact, such a ferromagnetic fluctuation has been detected in inelastic neutron scattering~\cite{20} and in Na-NMR measurements~\cite{Lang,21}.

Yoshizumi \textit{et al}. proposed that the difference between the CW and PP phases is related to this electron pocket~\cite{15}. Based on the rigid band picture depicted in Fig. 1, the Fermi level at $E_{\mathrm{F}}$ lies below the bottom of the dip in the PP phase for $x$ $<$ $x_{\mathrm{b}}$, touches the bottom at $x$ = $x_{\mathrm{b}}$ (the Lifshitz critical point), and becomes higher than the bottom to generate a small electron pocket around the $\Gamma$ point for $x$ $>$ $x_{\mathrm{b}}$, resulting in the CW phase. Thus, one would expect that a pocket-vanishing type LT would be achieved by band-filling control if one could control the Na content precisely across $x$*
in Na$_x$CoO$_2$.
Furthermore, it would be interesting to test whether such a simple LT on the basis of the rigid band picture as mentioned above occurs in such a strongly correlated electron system. 

\begin{figure}
\includegraphics[width=6.3cm]{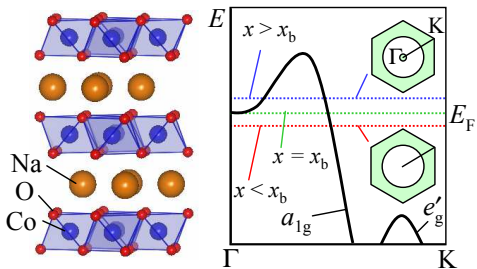}
\caption{\label{fig1}(color online) Schematic drawings of the crystal structure (left) and the band dispersion along the $\Gamma$-K line (right) of Na$_x$CoO$_2$. The Fermi level touches the bottom of the dip of the $a_{\mathrm{1g}}$ band at the $\Gamma$ point when $x$ = $x_\mathrm{b}$. A small electron pocket appears 
for $x$ $>$ $x_\mathrm{b}$, as depicted in the inset~\cite{15}.
}

\end{figure}

\begin{figure}
\includegraphics[width=7.5cm]{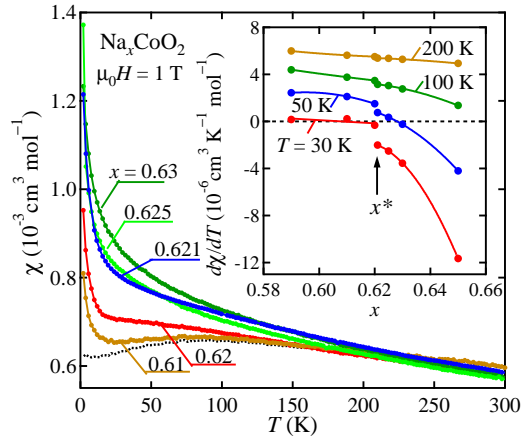}
\caption{\label{fig2}(color online) 
Temperature dependence of the magnetic susceptibility $\chi$ measured on cooling at a magnetic field of 1 T for a series of polycrystalline samples of Na$_x$CoO$_2$ for 0.61 $\le$ $x$ $\le$ 0.63. 
The dotted curve represents data for $x$ = 0.61 after subtraction of Curie-like contribution
from 0.2\% impurity spins.
The inset shows the $x$ dependence of the temperature derivative of $\chi$
at various temperatures, where curves are shifted upward by 2, 4, and 6 $\times$ 10$^{-6}$ cm$^3$ K$^{-1}$ mol$^{-1}$ for 50, 100 and 200 K data, respectively.
}

\end{figure}

A series of polycrystalline samples of Na$_x$CoO$_2$ with 0.55 $\le$ $x$ $\le$ 0.72 was prepared using a method modified from that used by Yoshizumi \textit{et al}~\cite{15}. First, Na$_{0.72}$CoO$_2$ was synthesized by a solid-state reaction.
Na$_{0.50}$CoO$_2$ was then obtained by stirring the product in a CH$_3$CN solution together with an excess amount of I$_2$. Samples with $x$ = 0.55, 0.59, 0.61, 0.62, 0.63, and 0.65 were prepared by mixing and reacting Na$_{0.50}$CoO$_2$ and Na$_{0.72}$CoO$_2$ in appropriate ratios in an evacuated quartz tube at 200$^{\circ}$C for 24 h~\cite{240}; this temperature is high enough to promote Na diffusion and prevent Na ordering. To achieve an even more precise control of $x$, a similar procedure was followed; samples with $x$ = 0.621 and 0.625 were obtained by reacting Na$_{0.62}$CoO$_2$ and Na$_{0.63}$CoO$_2$ in the same manner. All the samples were quenched from 200$^{\circ}$C to room temperature to prevent Na ordering, the absence of which was confirmed by the absence of an observed superstructure in electron diffraction experiments. The Na content $x$ was confirmed to be identical to the nominal value by chemical analysis using the inductively-coupled plasma method. Although the experimental resolution was limited to $\pm$0.01, the relative variation of $x$ should be guaranteed down to 0.001 or even less with the technique described here.
In fact, we checked a systematic variation in $x$ by observing a linear change in the lattice constants as a function of $x$, as determined by powder X-ray diffraction experiments. 

\begin{figure}
\includegraphics[width=7.3cm]{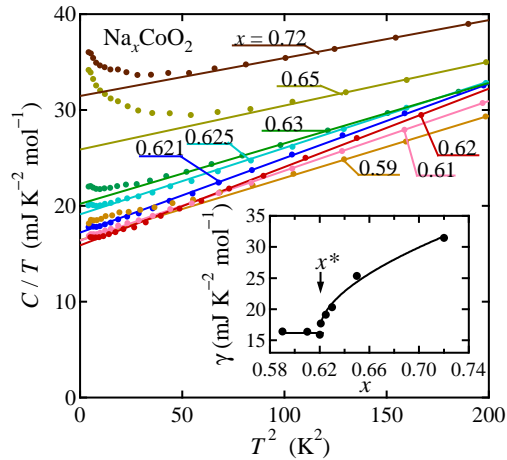}
\caption{\label{fig3}(color online) Heat capacity divided by temperature $C$ / $T$ versus $T^2$ plot for Na$_x$CoO$_2$ for 0.59 $\le$ $x$ $\le$ 0.72. A line on each data set is fitted to the form $C$ / $T$ = $\gamma$ + $\beta _3 T^2$
with appropriate $\beta _3$ values of 0.04-0.08 mJ K$^{-4}$ mol$^{-1}$.
The inset shows the $x$-dependence of $\gamma$ as deduced from the fit.
}
\end{figure}

The temperature dependence of the magnetic susceptibility $\chi$ changes its behavior abruptly across a critical Na content $x$*. Figure 2 shows selected $\chi$ data near $x$*.
All the curves merge into a single curve above 200 K, but exhibit two kinds of behavior at low temperatures: the $\chi$ curves for $x$ = 0.61 and 0.62 become nearly temperature independent
below $\sim$100 K or show decrease on cooling, 
while those for $x$ = 0.621, 0.625, and 0.63 show monotonic increases on cooling, following the CW law. Another Curie-like increase is observed below 10 K for each sample, which can be ascribed to impurity spins corresponding to approximately 0.5\% of all Co atoms. Therefore, the ground state of $x$ $\le$ 0.620 is basically a PP metal, while that of $x$ $\ge$ 0.621 is a CW metal. Surprisingly, the critical content $x$* is located in a very narrow range of 0.620 $<$ $x$* $<$ 0.621~\cite{24}. 

The dramatic change at $x$* is clearly demonstrated in the temperature derivative of $\chi$ shown in the inset of Fig.~2. $d \chi$/$dT$ at 30 K is nearly equal to 0 for $x$ $\le$ 0.620, suddenly drops at $x$ = 0.621, and decreases gradually with increasing $x$ for $x$ $\ge$ 0.621. A sudden drop at $x$* is also observed at 50 K, 
indicating that the transition is sharp even at considerably high temperatures. 

An abrupt change at $x$* is also observed in the heat capacity $C$ at low temperatures. The Sommerfeld coefficient $\gamma$, which is proportional to the density of states (DOS) at the Fermi level, is determined by fitting the data to the form $C$ / $T$ = $\gamma + \beta _3T^2$, as shown in Fig. 3.
There is an unexpected upturn in $C$ / $T$ at low temperature, especially for $x$ = 0.65 and 0.72, which may be ascribed to a Schottky contribution corresponding to a small entropy less than 1\% of $R$ln2, probably coming from localized electrons. Thus determined $\gamma$ is almost constant at $\sim$ 16 mJ K$^{-2}$ mol$^{-1}$ for $x$ $\le$ 0.620 and increases suddenly at $x$ = 0.621, followed by a gradual increase with increasing $x$ indicative of 0.620 $<$ $x$* $<$ 0.621 (inset of Fig. 3), in good agreement with the above results from $\chi$. The constant $\gamma$ for $x$ $<$ $x$* means that the added electrons occupy the large hole FS with the two-dimensional character, and the increase in $\gamma$ for $x$ $>$ $x$* implies that an additional FS, that must be an electron pocket around the $\Gamma$ point, emerges there. 
By fitting the $x$ dependence of $\gamma$ in the CW phase (0.621 $\le$ $x$ $\le$ 0.72) to the form of $\gamma$ = $\gamma$ ($x$ = 0.620) + $\alpha$($x - x_{\mathrm{b}}$)$^\beta$, it was determined that $\alpha$ = 0.050(8) J K$^{-2}$ mol$^{-1}$, $\beta$ = 0.49(6), and $x_{\mathrm{b}}$ = 0.619(2),
suggesting that
the electron pocket possesses a three-dimensional character, as found by the band-structure calculations~\cite{16,17,18}. 

\begin{figure}
\includegraphics[width=7.5cm]{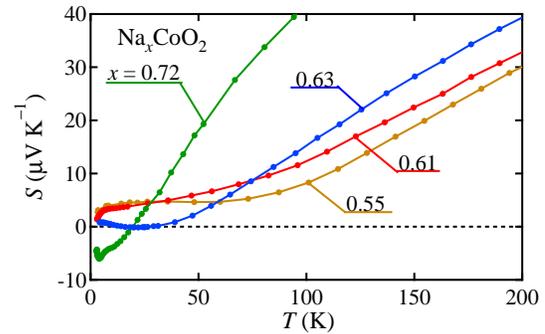}
\caption{\label{fig4}(color online) Temperature dependence of the thermoelectric power of Na$_x$CoO$_2$ polycrystalline samples.
}
\end{figure}

Thermoelectric power $S$ is a physical quantity sensitive to a change in the topology of a FS. As shown in Fig. 4, the $S$ curves of $x$ = 0.55 and 0.61 exhibit similar $T$-dependences with positive values, 
which is consistent with a FS consisting of a single hole sheet for $x$ $<$ $x$*. In contrast, the $S$ curve for $x$ = 0.63 is almost 0 below 50 K, 
and 
the $S$ curve for $x$ = 0.72 changes its sign at 20 K. These results indicate that both positive and negative carriers exist for $x$ $>$ $x$*, implying the emergence of an electron pocket. 
In addition, $x$-dependences previously reported for the Hall coefficient and for resistivity~\cite{13,15} are consistent with these results.
In summary, all these physical properties exhibit drastic changes at $x$*, indicating that a pocket-vanishing type LT occurs at $x$* between the PP and CW metal phases.
Note that the observed changes at $x$* would be irrelevant to the 
Na ordering~\cite{Na-order} and the complicated low-temperature magnetism present at $x$ $\ge$ 2/3~\cite{Schulze}, because $x$* is well below 2/3.

Thus far, many experimental results have been interpreted assuming the rigid band picture shown in Fig. 1. A conventional LT should be continuous, so that one would expect no discontinuities for thermodynamic quantities at $T$ = 0. However, there are obviously more drastic changes at $x$* in Na$_x$CoO$_2$. More importantly, a conventional LT should occur as a crossover at elevated temperatures. Consider, for example, the case of $x$ = 0.620, which is just less than $x$* by at most 0.001. Given a bandwidth of 0.7 eV for the two-dimensional $a_{\mathrm{1g}}$ band based on the band-structure calculations, a difference in the band filling by $\Delta x$ = 0.001 corresponds to a difference in $E_{\mathrm{F}}$ of only 4 K. Thus, one would observe a clear difference between the two phases at low temperature well below 4 K and a crossover at higher temperatures, because a number of electrons which have been thermally-excited from the large hole sheet should exist in the electron pocket at higher temperatures.

\begin{figure}
\includegraphics[width=6.5cm]{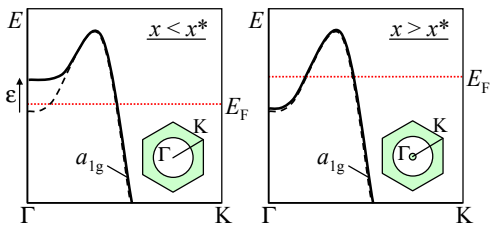}
\caption{\label{fig5}(color online) 
Schematic drawings of the band dispersion along the $\Gamma$-K line for $x$ $<$ $x$* (left) and $x$ $>$ $x$* (right), when the instability of the small FS around the $\Gamma$ point is considered. Thick solid and broken lines designate dispersions with and without the instability respectively. 
}
\end{figure}

Obviously, this is not the case observed for the $\chi$ shown in Fig. 2. The $\chi$ curves for $x$ = 0.620 and 0.621 merge into a single CW curve at high temperatures, while remaining separate below 200 K. Moreover, a discontinuous jump in $d\chi$/$dT$ is observed below 100 K. These results indicate that the crossover temperature $T_{\mathrm{co}}$ is $\sim$100 K, which is much higher than 4 K expected from $\Delta$$x$ = 0.001. Therefore, we think that an essentially different mechanism is operating for the PP-CW transition in Na$_x$CoO$_2$. 

The discontinuities and high $T_{\mathrm{co}}$ values observed for the phase transition of Na$_x$CoO$_2$ can be qualitatively understood by taking into account the instability of the small electron pocket around the $\Gamma$ point. In the compound under study, electron correlations with ferromagnetic fluctuations tend to be switched on as a small FS appears, as evidenced by the sudden appearance of CW magnetism. The reason for this is not clear, but possibly related to a flat band near $E_{\mathrm{F}}$ that favors ferromagnetic interactions~\cite{25}. Therefore, it is likely that electrons in the dip would cause an enhancement of the total energy by a certain energy $\varepsilon$, compared with the case where all electrons occupy the outer hole sheet. As schematically depicted in the left panel of Fig. 5, even if $x$ is slightly larger than $x_{\mathrm{b}}$ in the rigid band picture, electrons should occupy only the outer hole sheet, because the bottom of the dip has been pushed up by $\varepsilon$ to avoid occupying the small electron pocket. When the $E_{\mathrm{F}}$ is further elevated with increasing $x$, however, such a large band deformation is no longer sustainable, and the dip must suddenly drop to generate an electron pocket, as shown in the right panel of Fig. 5. In this scenario, one would expect the PP-CW transition to occur discontinuously even at finite temperatures of $T$ $<$ $\varepsilon$ / $k_{\mathrm{B}}$. 

This mechanism is reminiscent of the first-order Mott transition caused by competition between Coulomb interactions and the screening effect. 
In addition, it has been pointed out that Lifshitz transition can become discontinuous in the presence of strong electron correlation~\cite{2}.
By analogy with the Mott transition, the first-order phase-transition line of the PP-CW transition in Na$_x$CoO$_2$ should have a critical end point in the $x$-$T$ phase diagram, which may be located at $x$ $\sim$ $x$* and $T$ $\sim$ 100 K. It would be interesting to be able to detect a phase transition as a function of $T$.
However, it seems difficult to do so in actual experiments, because the phase line may be almost parallel to the $T$-axis in the phase diagram. 
As an alternative, it may be possible to observe a transition by applying a magnetic field or a pressure; the latter possibility is particularly intriguing, because it may effectively change the energy of the $a_{\mathrm{1g}}$ band by distorting the CoO$_6$ octahedra~\cite{26}. 


In conclusion, this research has examined the PP-CW transition of Na$_x$CoO$_2$ using band-filling control and has found a discontinuous LT of the pocket-vanishing type at $x$* between 0.620 and 0.621. A mechanism has been proposed which explains various experimental results consistently, assuming the instability of a small FS.
Na$_x$CoO$_2$ is a fascinating compound, not only in terms of its thermoelectric or superconducting properties, but also because of its unique electronic phase transition.

We thank M. Ichihara for electron diffraction measurements and Y. Kiuchi for chemical analyses.  We are also grateful to M. Imada, H. Harima, R. Arita, M. Ogata, and M. Mochizuki for stimulating discussion. This work was partly supported by a Grant-in-Aid for Scientific Research on Priority Areas ``Novel States of Matter Induced by Frustration'' (No. 19052003).

\end{document}